\documentclass[aps, prb, superscriptaddress, twocolumn, amsfonts, amsmath, amssymb, floatfix]{revtex4-2}
\usepackage{graphicx}
\usepackage{float}
\usepackage{sidecap}
\usepackage{dcolumn} 
\usepackage{bm}
\usepackage{epsfig}
\usepackage{braket}
\usepackage{color}
\usepackage{ulem}
\usepackage{amsmath}
\usepackage{caption}
\usepackage{subcaption}

\usepackage{ragged2e}
\DeclareCaptionFormat{justified}{\justifying#1#2#3}
\usepackage[toc,page]{appendix}
\usepackage[colorlinks=true, letterpaper=true, pdfstartview=FitV, linkcolor=blue, citecolor=blue, urlcolor=blue]{hyperref}

\begin{document}
\title{Tunable Valley Filtering and Splitting via Gate Control in Wrinkled Graphene}
\author{C. I. Osuala }
\email{cosuala@stevens.edu}
\affiliation{Department of Physics, Stevens Institute of Technology, Hoboken, NJ 07030 USA.}
\author{A. Belayadi}

\affiliation{Department of Physics, University of Science and Technology Houari Boumediene, Bab-Ezzouar, Algeria.}

\author{M. Moutan}
\affiliation{Department of Physics, King Fahd University of Petroleum and Minerals, 31261 Dhahran, Saudi Arabia.}
\author{S. Elatresh}
\affiliation{Department of Physics, King Fahd University of Petroleum and Minerals, 31261 Dhahran, Saudi Arabia.}
\affiliation{Advanced Quantum Computing Center,
King Fahd University of Petroleum and Minerals, 31261 Dhahran, Saudi Arabia.}
\author{A. Johanson}
\affiliation{Institut für Physik, Martin-Luther-Universität Halle-Wittenberg, D-06099 Halle (Saale), Germany and Halle-Berlin-Regensburg Cluster of Excellence CCE, Halle, Germany}

\author{A. Abbout}
\email{adel.abbout@kfupm.edu.sa }
\affiliation{Department of Physics, King Fahd University of Petroleum and Minerals, 31261 Dhahran, Saudi Arabia.}
\affiliation{Advanced Quantum Computing Center,
King Fahd University of Petroleum and Minerals, 31261 Dhahran, Saudi Arabia.}

\begin{abstract}
We investigate the use of strain-induced wrinkles in graphene nanoribbons to achieve tunable valley filtering and splitting, crucial for advancing valleytronic devices.  By combining tight-binding simulations, strain tensor analysis, and quantum transport calculations, we explicitly demonstrate that wrinkle-induced pseudomagnetic fields facilitate current splitting and valley-selective filtering in folded graphene nanoribbons based on wrinkle parameters like height, strain spreading/deformation and wrinkle orientation (tilt angle). 
Moreover, we show that the application of a top-gate voltage provides dynamic control over valley splitting, allowing a continuous transition from well-separated valley currents to valley merging as the gate voltage increases. In other parameter regimes, valley filtering arises from an initial current-splitting mechanism, exhibiting a clear gate-controlled transition between distinct transport behaviors. Our results offer new insights into strain-engineered valley transport and suggest a promising pathway toward practical graphene-based valleytronic devices.

\end{abstract}
\maketitle
\section{\label{sec1} Introduction}
Graphene has emerged as a testbed for studies of electronic and transport properties due to its unique band structure, high carrier mobility and linear dispersion relation near the Dirac points \cite{castro2009electronic, das2011electronic, goerbig2011electronic,peres2010colloquium}. Most intriguing electronic properties of graphene arise from its Dirac-like massless charge carriers around the $K$ and $K^{'}$ valleys of the Brillouin zone, giving rise to a valley degree of freedom.
Recent studies have shown that, in addition to charge and spin, electrons in solids also possess a valley pseudospin that can be utilized to create the next generation of electronic devices in emerging field valleytronics \cite{hadadi2023pseudo, settnes2016graphene}.  Valleytronics aims to manipulate and control valley-polarized currents, providing new opportunities for information processing and quantum computing applications.

Despite extensive theoretical and experimental efforts, realizing practical valleytronic devices remains a significant challenge. The generation, control, and detection of valley-polarized currents require precise engineering of graphene’s electronic structure.  Valley polarization was first experimentally demonstrated in $\text{MoS}_2$ monolayer using circularly polarized light for optical excitation \cite{zeng2012valley}. This method takes advantage of the distinct optical selection rules that apply differently to $K$ and $K^{'}$ valleys, enabling valley-specific transitions \cite{zeng2012valley,jones2013optical,cao2012valley,mak2012control}. However, achieving electrically tunable valley filters in graphene has proved more difficult due to the intrinsic symmetry of the material and the lack of an intrinsic band gap. Several mechanisms have been proposed to enable valley filtering, including the use of perfect zigzag nanoribbons \cite{rycerz2007valley}, valley Hall effects \cite{xiao2007valley,gorbachev2014detecting,shimazaki2015generation,sui2015gate}, trigonal warping \cite{garcia2008fully,gold2021coherent}, valley-dependent quantized states, electron scattering at line defects \cite{gunlycke2011graphene,xiao2012valley}, and external fields \cite{abergel2009generation}. Nonetheless, many of these approaches require complex and specialized experimental setups or external magnetic fields, limiting their practical implementation.

A promising alternative for valley manipulation in graphene involves strain engineering \cite{amorim2016novel,zhai2019electron}. 
The realization of strain engineering has become a crucial tool in tailoring electronic properties, as local deformations can induce pseudomagnetic fields (PMF), modify valley-dependent transport, and enable novel device functionalities. Strain-induced deformations, such as wrinkles and folds \cite{tomori2011,reserbat2014strain, pacakova2017,jiang2017,meng2013hierarchy,lim2015structurally,faria2020valley}, introduce out-of-plane distortions which locally alter the electronic structure, leading to spatially varying gauge fields that affect charge carriers differently at the $K$ and $K^{'}$ valleys. These gauge fields act as pseudomagnetic fields, selectively affecting electrons in the $K$ and $K^{'}$ valleys and leading to valley-dependent transport properties. 

A critical consequence of mechanical deformation in graphene-based systems is the emergence of these pseudomagnetic fields (PMFs), which are induced by non-uniform strain through the strain tensor’s coupling to the electronic structure. It arises from the formal analogy between strain-induced gauge fields and electromagnetic vector potentials, where spatial variations in the strain tensor $\epsilon_{ij}$ generate a pseudovector potential $A$. The resulting pseudomagnetic field $B=\nabla \times A$ alters the charge carrier dynamics without breaking time-reversal symmetry. In folded graphene nano-ribons (GNRs), curvature and anisotropic strain distributions create spatially varying $A$ profiles, leading to pseudomagnetic fields that polarize electronic states, modify transmission pathways, and split degenerate energy levels. These fields are particularly pronounced in regions of high curvature or strain gradients, such as the folded edges of a nanoribbon, where the interplay between geometric deformation and sublattice symmetry breaking generates valley-dependent scattering.

The presence of pseudomagnetic fields has been experimentally observed in strained graphene, with reported field strengths reaching hundreds of teslas in highly deformed regions. This phenomenon provides an exciting pathway for controlling valley polarization without requiring external magnetic fields, making strain a powerful tool for valleytronic applications. Recent studies have demonstrated that these out-of-plane deformations in graphene nanoribbons can lead to phenomena such as valley polarization,  valley filtering, and valley splitting \cite{stegmann2016current,stegmann2018current}. However, a major challenge in practical valleytronic devices remains the lack of tunability and dynamic control over valley-dependent responses.

In this work, we investigate the effects of wrinkles strain induced out-of plane deformations on valley-polarized transport in graphene nanoribbon. We demonstrate that such strain-engineered pseudomagnetic fields act as a powerful tool for current splitting and energy-selective filtering in folded GNRs. By combining tight-binding simulations, strain tensor analysis, and quantum transport calculations, we show that the pseudomagnetic field $B$ spatially separates charge carriers of opposite momenta while suppressing backscattering in specific energy windows. We demonstrate how
these structural modifications generate pseudomagnetic fields that lead to valley current splitting, filtering, and trapped wave-like behavior. We explore how top gate modulation enables tunable valley filtering and valley splitting. By applying a top gate voltage $V_g$ on a system with a fixed wrinkle deformation, we demonstrated that Valley splitting can be dynamically tuned, and in some regimes, transitions into valley merging as $V_g$ increases. In other regimes, valley filtering behavior emerges from an initial current splitting, showcasing a clear transition controlled by gate voltage.
Our findings provide new insights into strain-engineered valley transport and offer a potential route toward the practical realization of graphene-based valleytronic devices. By bridging the gap between theoretical models and experimental feasibility, this study contributes to advancing the design of strain-controlled valley filters and paves the way for future valleytronic applications.
\section{\label{sec2} Strain Theoretical Model}
\begin{figure}[h]
	\includegraphics[width=0.45\textwidth]{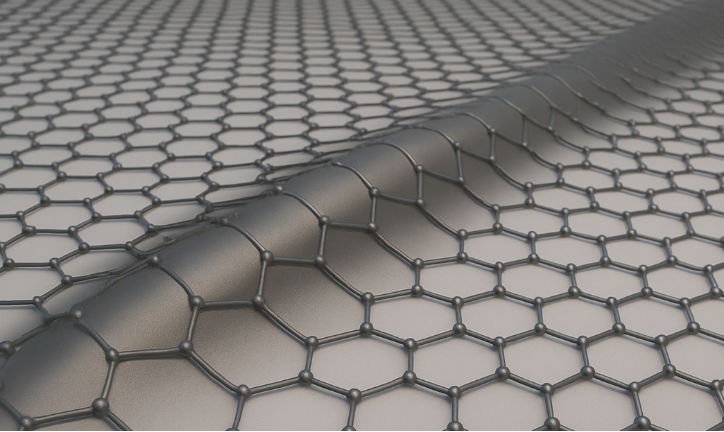}
	\caption{Schematic illustration of a strain-induced wrinkle in a graphene nanoribbon. The one-dimensional out-of-plane deformation extends along the z-direction and spans the full width of the nanoribbon. The structure may be placed on a substrate, allowing electrostatic gate voltages to be applied for tunable transport control.}
	\label{fig:sys}
\end{figure}

Fig.~\ref{fig:sys} shows a schematic of a flat sheet of graphene, subjected to a one dimensional out-of-plane deformation refered to as wrinkle in the text. This ridge-like deformation can be the result of a compressive strain appearing upon cooling graphene after high temperature growth or annealing \cite{PhysRevX,Li2009}. Indeed, the inextensible but bendable nature of the carbon bond favors the transition from a complete adhesion to a substrate to a delamination \cite{Bao2009,Scharf}. The typical width and maximum height of the deformation are about few nanometers. Four leads are attached to the system (see Fig.~\ref{fig:T(E)}) to investigate charge transport through the wrinkle. The injection lead is intentionally made narrow in order to focus the incident current onto the wrinkle region, while the remaining three leads are considerably wider to efficiently collect the scattered charge carriers and ensure complete current detection. The nanoribbon is described by the nearest-neighbor tight-binding Hamiltonian

\begin{equation}
	H= \sum_{m} eV(z_m)\Big(a_{m}^{\dag}a_{m} + b_{m}^{\dag}b_{m}\Big)+\sum_{\langle m,n\rangle} t_{mn}a_{m}^{\dag}b_{n} + \text{h.c.}
\end{equation}
where the $\sum_{\langle m,n\rangle}$ runs over the nearest neighboring sites on sublattices A and B. $t$ is the hopping parameter which varies in space due to a change in the interatomic distances when a lattice deformation is present. A voltage gate $V$ is applied to modulate the carrier density  in the wrinkle. Indeed, this voltage is the result of a uniform electric field in the $z$-direction  that keeps all the sites in the same level at equal potential. With a proper gauge, the potential on the flat graphene sheet is brought to zero while on the deformation it varies as a function of the height $z_m$ of the correponding site. With this, we can write $V(z)=V_g\frac{z}{h}$, where $V_g$ represents the maximum voltage at the maximum height of the wrinkle. The hopping parameter due to strain is modified as $t_{mn} \approx t_{0}e^{-\beta(d_{mn}/a_{0}-1)}$ \cite{PhysRevB.94.125422, PhysRevB.80.045401},
where $a_{0}=1.42 \mathrm{~\r{A}}$, $\beta=3.37$, $t_0 = -2.7 \mathrm{~eV}$, and $d_{mn}$ is the modified bond length of the strained lattice vector expressed as \cite{vozmediano2010gauge}

\begin{equation}
	d_{mn} = \frac{1}{a_0}(a_o^2+\epsilon_{xx}x_{mn}^2+\epsilon_{yy}y_{mn}^2+2\epsilon_{xy}x_{mn}y_{mn})  
\end{equation}
where the strain tensor is classically expressed as \cite{landau2013course} 
\begin{equation}\label{eq3}
	\epsilon_{mn}=\frac{1}{2}(\partial_n u_m + \partial_m u_n+ \partial_m z\partial_n z), 
	\hspace{15px} m,n=x,y  
\end{equation}
$u_m$, $u_n$, represent the $x$-$y$ displacement. In our case, we assume that the deformation doesn't induce an inplane displacement of the sites and therefore the model for the wrinkle in the graphene nanoribbon,  Eq.\ref{eq3} reduces to

\begin{equation}
	\epsilon_{mn}=\frac{1}{2} \partial_m z\partial_n z;
\end{equation}

we consider a Gaussian wrinkle-shaped deformation \cite{Nancy1}
described by the following expression
\begin{equation}\label{eq5}
	z(x,y)=h\exp(-\frac{r^2}{2\sigma^2})
\end{equation}
with 
$r=x\cos(\theta) + y \sin(\theta)$. Here $h$ denotes the maximum deformation height and $\sigma$ characterizes its spacial extent. The angle $\theta$ is introduced to account for tilted wrinkles, allowing the wrinkle main axis to be oriented at arbitrary angle with respect to the transport direction.  It is worth noting that the expression in Eq. \ref{eq5} differs from the commonly studied two-dimensional Gaussian deformation (nanobubble) that is spatially localized \cite{hadadi2023pseudo,Jack,Nancy1}. Instead, it corresponds to a one-dimensional deformation, as illustrated schematically in Fig.~\ref{fig:sys}. In strained graphene, the resulting deformation gives rise to a valley-dependent pseudo–magnetic field, which enters the low-energy Dirac Hamiltonian as \cite{castro2009electronic}
\begin{equation}\label{eq6}
	H_k =v_{F}\bm{\sigma} \cdot \bf{p} \pm e \bf{A}
\end{equation}
where $\bf p$ is the momentum measured from $\bf k$ point, $\bm{\sigma}=[\sigma_x,\sigma_y]$ (Pauli matrices), $v_{F}$ is the Fermi velocity and $\pm$ corresponds to the $K$ or $K^{'}$ valley. 
Notably, in this low-energy model, there is no inherent coupling between the two valleys, so the Hamiltonian for the $K^{'}$ valley is obtained by applying the transformations
$\bm{\sigma} \rightarrow-\bm{\sigma}$ and $\bf{A}\rightarrow \bf{A}$ such that time reversal symmetry is conserved. 

This valley-dependent sign difference, combined with the spatial variation of the PMF $\bf{B=\nabla\times \bf{A}}$,  forms the basis of our approach to manipulate the valley degree of freedom
using strain engineering.

The vector potential $\bf A$ in Eq.\ref{eq6} is in the form of two-dimensional strain tensor $\epsilon_{ij}(x,y)$ \cite{jones2014designing,suzuura2002phonons};
\begin{equation}
	A=-\frac{\hbar\beta}{2ea_{o}} \begin{bmatrix}
		\epsilon_{xx}-\epsilon_{yy}\\
		-2\epsilon_{xy}
	\end{bmatrix}
\end{equation}
and its corresponding out-of-plane PMF $\bf B$ reads,

\begin{equation}
	B = -\frac{\hbar\beta rz^{2}}{ea_{o}\sigma^4}\bigg(1-\frac{r^2}{\sigma^2}\bigg)\sin{3\theta} 
\end{equation}
The PMF changes sign at $r=\sigma$ and becomes exponentially smaller for larger distances $r$. It reaches its maximum value for $\theta=n\frac{\pi}{6}, \,\text{for } n \text{ odd}$
and becomes zero everywhere for $\theta=n\frac{\pi}{6}, \, \text{with }n\text{ even}.$


\begin{figure}[h]
	\centering
	\includegraphics[width=0.45\textwidth]{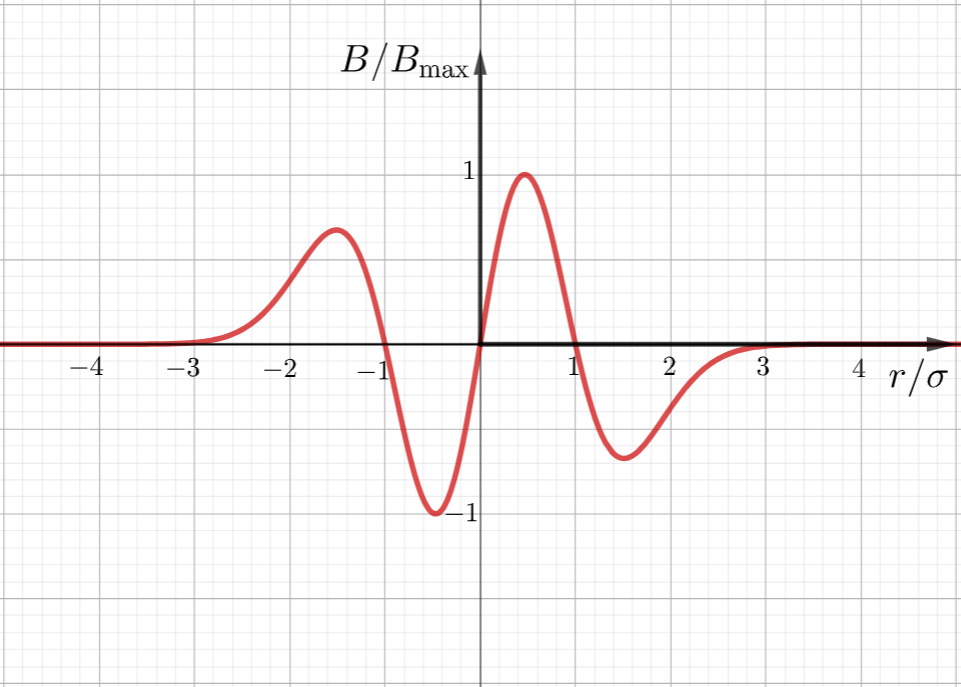}
	\caption{Typical  out-of-plane pseudo-magnetic field as a function of the position from the wrinkle ($\theta=\pi/2$). The PMF changes sign and might exibit maximum magnitudes of hundreds of Tesla. The PMF becomes negligible for $|r|>3\sigma$}
	\label{fig:B(r)}
\end{figure}
Fig. \ref{fig:B(r)} shows the PMF spatial profile across  the GNR wrinkle. This PMF plays a key role in modulating valley-dependent electronic transport \cite{hadadi2023pseudo, Hadadi2025,stegmann2016current}.
This bipolar structure is characteristic of a local wrinkle deformation, which causes the contributions of opposite PMF on either side of the central peak to  rapidly decay toward zero away from the center, becoming negligible for $r > 3\sigma $. This localization implies that the wrinkle-induced pseudo-magnetic effects are spatially confined, making them suitable for local valley manipulation without affecting the entire system. Also, this spatial variation in the field breaks sublattice symmetry differently for $K$ and $K^{'}$ valleys, giving rise to valley splitting and filtering effects depending on the angle of incidence  of electrons. The field polarity effectively acts like a valley-dependent magnetic barrier, allowing selective control of valley currents via wrinkle engineering. 

\begin{figure}[h]
\centering
\includegraphics[width=0.45\textwidth]{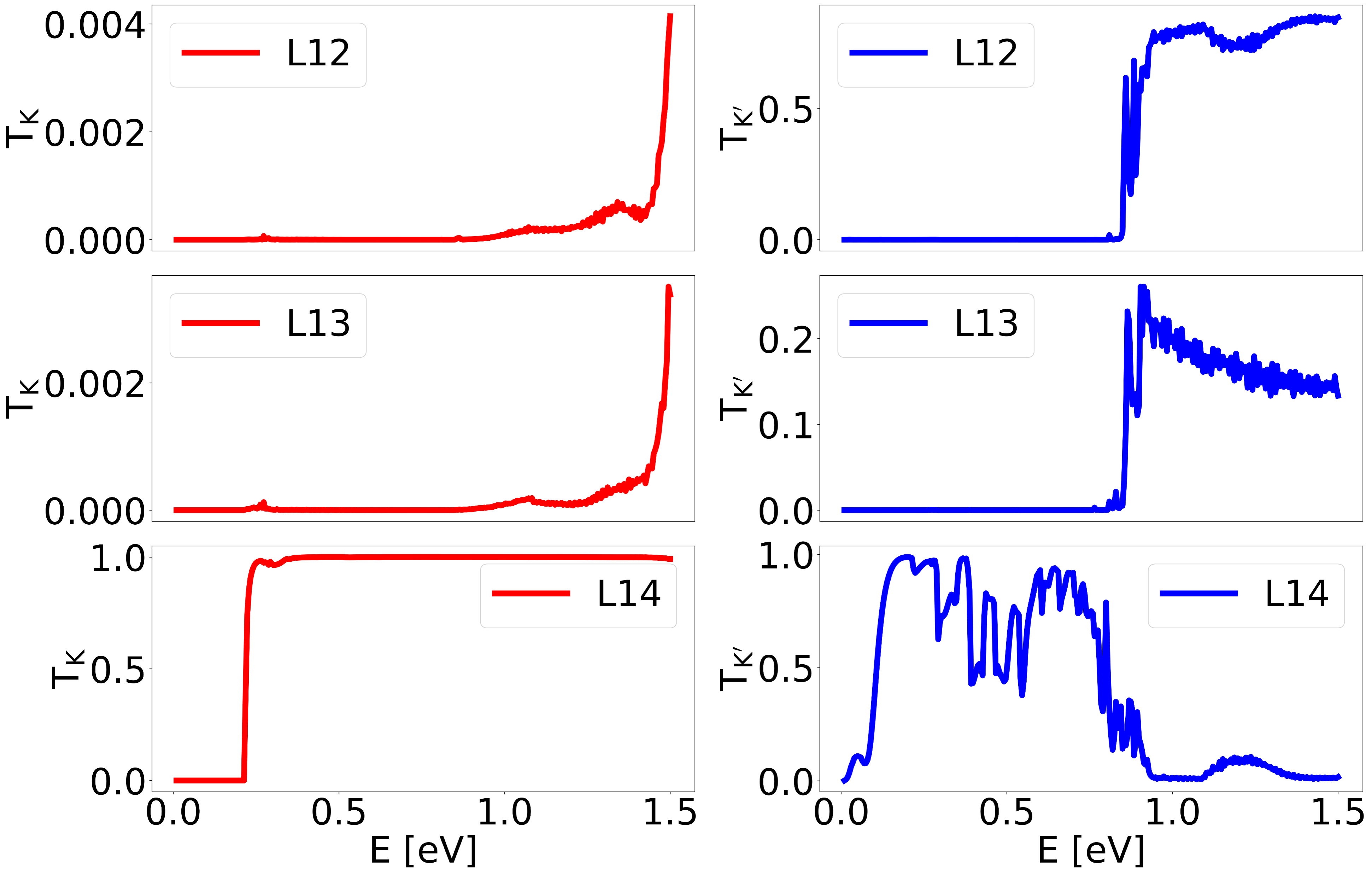}
\caption{Resolved transmission as a function of energy for $K$  (red) valley and $K'$ (blue) valley, for $h=8$ nm, $\sigma=4.65$ nm, $\theta=0.9 \text{ rad}$}
\label{fig:T(E)}
\end{figure}


\begin{figure}[h]
\centering
\begin{subfigure}[b]{0.45\textwidth}
\includegraphics[width=0.9\linewidth]{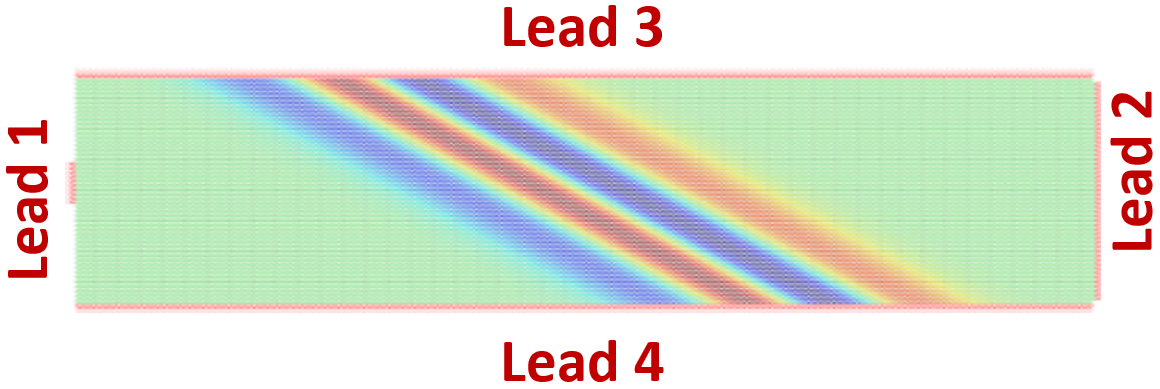}
\end{subfigure}
\begin{subfigure}[b]{0.45\textwidth}
\includegraphics[width=1\linewidth]{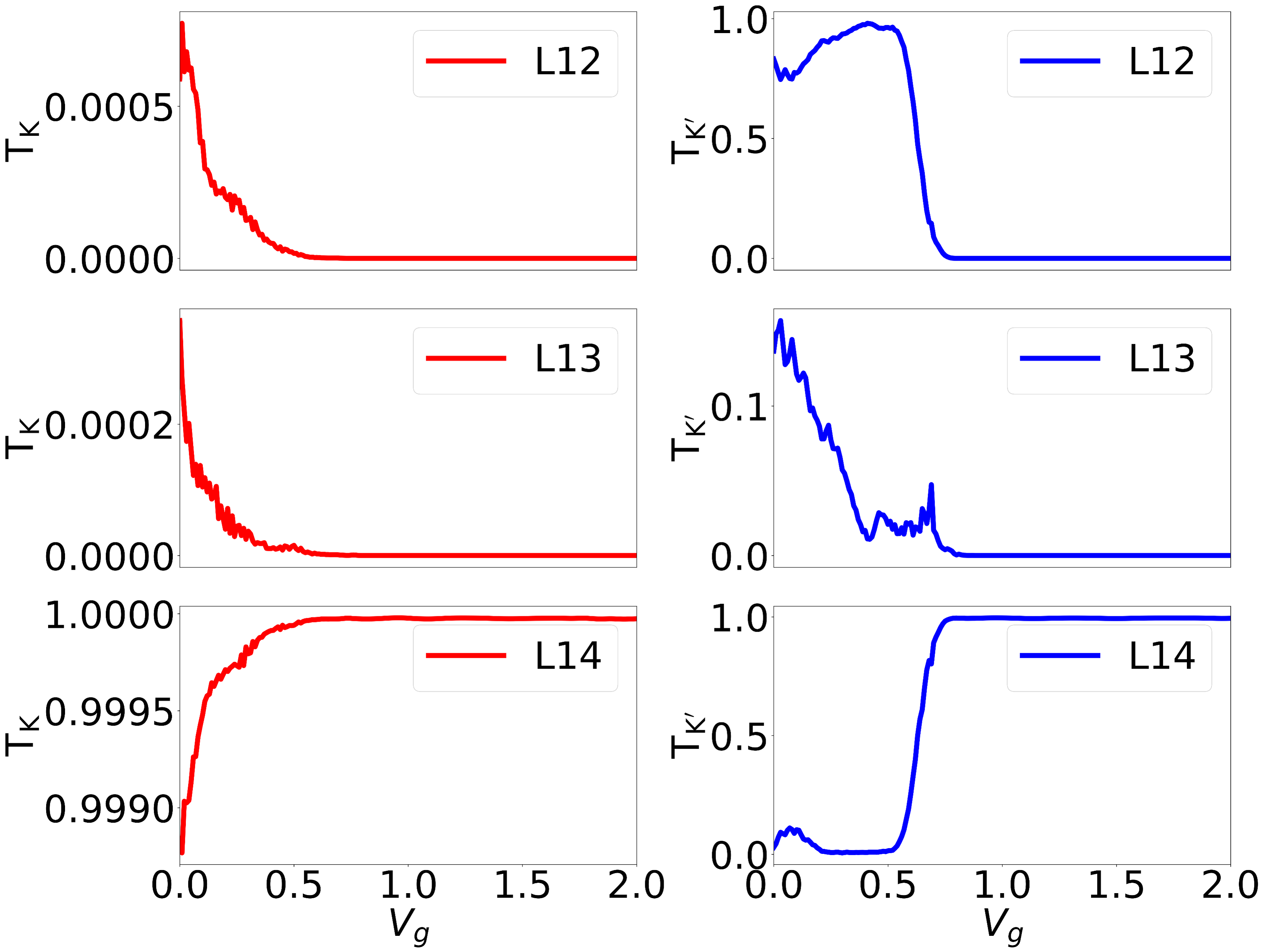}
\end{subfigure}	
\caption{(Top figure) Graphene connected to 4 leads represented by red sites at the edges (following Kwant convention). The current is injected from lead 1. The pseudo magnetic field for the tilted wrinkle is represented with red and blue colors for the two different signs. The graphene zigzag direction is along the x-direction. The curves in bottom graphs represent the resolved transmission as a function of gate voltage ($V_g$) for $K$  (red) valley and $K'$ (blue) valley, for $h=8$ nm, $\sigma=4.65\mathrm{~nm}$, $\theta=0.9 \text{ rad}$.}
\label{fig:T(Vg)}
\end{figure}
Fig.~\ref{fig:T(E)} presents the transmission probabilities of the two valleys into the different output leads. Here, $L{1 j}$ denotes the transmission from the injection lead $L1$ to lead $Lj$. The spatial arrangement of the leads is illustrated in Fig.~\ref{fig:T(Vg)}. From this figure, we observe that by appropriately selecting the energy, it is possible to achieve valley selective transport, such that only one valley is preferentially transmitted or deflected into a specific lead. Furthermore, we note that lead $L4$ collects the largest portion of the deflected current, which can be attributed to the tilt orientation of the wrinkle. 

Fig. \ref{fig:T(Vg)} shows the lead-resolved transmission probabilities for $K$ (red) and $K^{'}$ (blue) valley  as a function of gate voltage $V_g$. The electron is injected from the left lead L1, and the outgoing transmissions are analyzed at the right L2, top L3, and bottom L4 leads. For the $K$ valley electrons, we observe a strong transmission into the lead L4, with nearly unit probability over the entire range of $V_g$. By contrast, their transmission to the right and top leads remains negligible, indicating that the wrinkle combined with the gate efficiently directs $K$ valley carriers downward.
On the other hand, the $K^{'}$ valley electrons exhibit an opposite behavior where their transmission into the lead L4 is heavily suppressed, while significant transmission is initially observed especially at low $V_g$ into L3 and L2. However, as the $V_g$  increases, the transmission into both L2 and L3 decreases sharply, which signals to enhanced backscattering. 

We identify two distinct regimes of valley injection: splitting and filtering. This distinction is made  depending on where valley scattering occurs relative to the position of the wrinkle.

The splitting regime arises when an unpolarized injected current causes both valleys to be backscattered in the same region relative to the wrinkle, but at different angles. This angular separation results in a spatial splitting of the two valleys.

\begin{figure}[h]
\centering
\begin{subfigure}[b]{0.5\textwidth}
\includegraphics[width=1\linewidth]{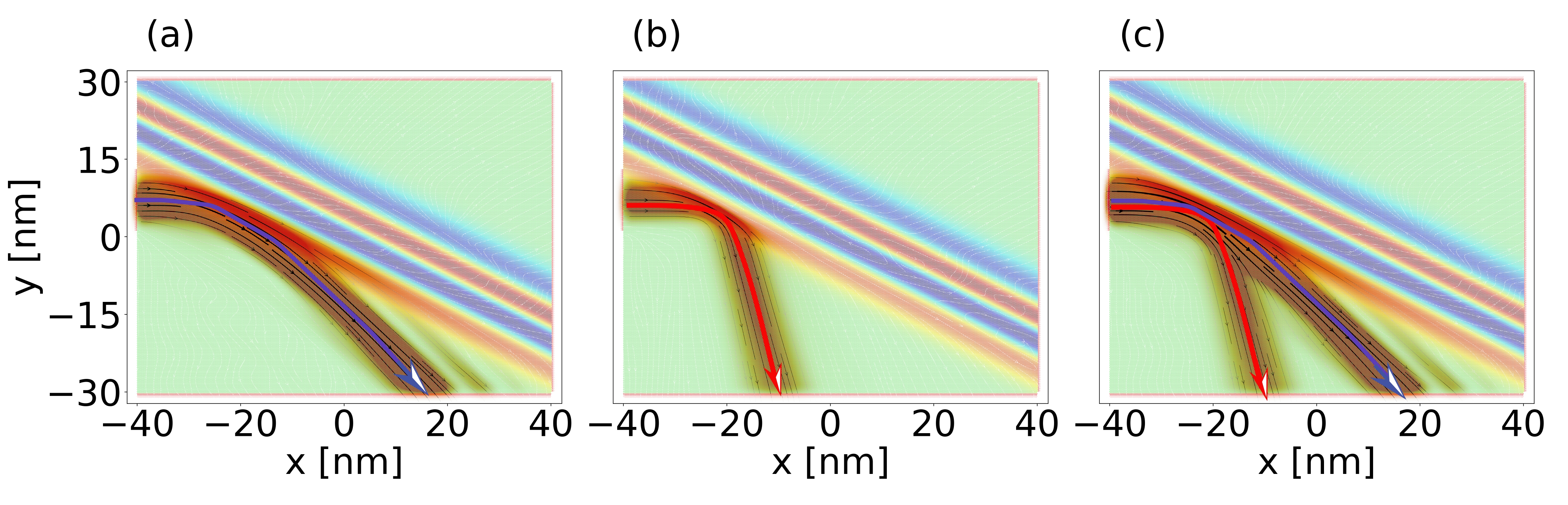}
\end{subfigure}
	
\begin{subfigure}[b]{0.5\textwidth}
\includegraphics[width=1\linewidth]{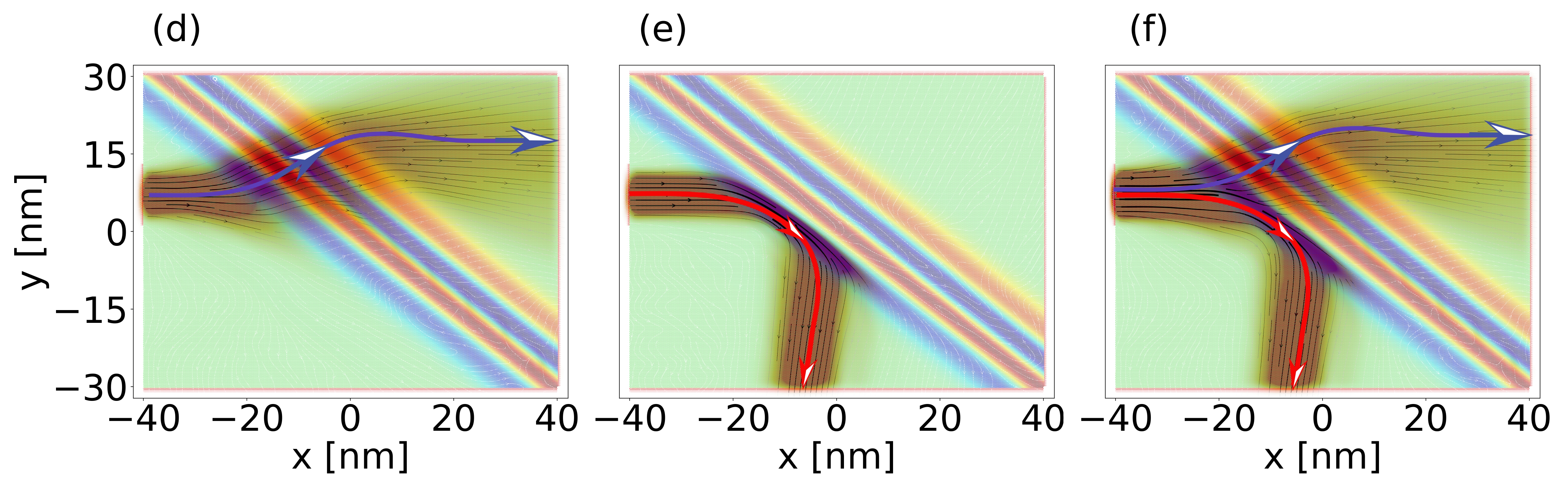}
\end{subfigure}	
\label{Figures5}
\caption{Current distribution for valley-polarized transport under valley-selective current injection (injection from lead 1 on the left side.). Panels (a-c)  illustrate the valley splitting regime for $h=7 \mathrm{~nm}, \sigma=4.65 \mathrm{~nm}$, and $\theta=1.1 \mathrm{~rad}$: (a) current injected from the $K$ valley, (b) current injected from the $K^{\prime}$ valley, and (c) resulting total current, showing spatial separation of the two valley contributions. Panels (d-f) correspond to the valley filtering regime for $h=8 \mathrm{~nm}, \sigma=4.65 \mathrm{~nm}$, and $\theta=0.90 \mathrm{~rad}$ : (d) current injected from the $K$ valley, (e) current injected from the $K^{\prime}$ valley, and (f) total current, demonstrating preferential transmission of a single valley. The blue/red arrows are  eye-guide for $K$/$K'$ valleys.
}
\label{fig5}
\end{figure}

The filtering regime, on the other hand, occurs when the wrinkle selectively backscatters only one valley, while allowing the other valley to transmit across the wrinkle. (See Fig.~\ref{fig5} for more details)

Fig.~\ref{fig5} (a-c) show the valley splitting regime for $h=7\mathrm{~nm}$, $\sigma=4.65$ nm, $\theta=1.10^{\circ}$. The charge carriers from $L1$ travel along the same path before interacting with the wrinkle. The wrinkle, however, induces a local strain that creates a valley-dependent pseudomagnetic field, and as a result, the $K$ valley current (Fig.~\ref{fig5}(a)) and $K^{'}$ valley current (Fig.~\ref{fig5}(b)) are deflected in the same side of the wrinkle but at differnt directions. The total current (unpolarized) carrying both valleys in Fig.~\ref{fig5}(c) exhibits well this valley splitting. 
Fig.\ref{fig5}(d-f) correspond to the valley filtering regime, achieved with $h=8$ nm, $\sigma=4.65$ nm, $\theta=0.90^{\circ}$. In this case, the angle of incidence are such that the wrinkle preferentially allows only one valley to pass. The $K$ valley current is transmitted through the wrinkle Fig.~\ref{fig5}(d), while the $K^{'}$ valley is reflected Fig.\ref{fig5}(e). The resulting total current Fig.~\ref{fig5}(f) confirms that the wrinkle acts as a valley filter.

\begin{figure}[h]
	\centering
	\includegraphics[width=0.5\textwidth]{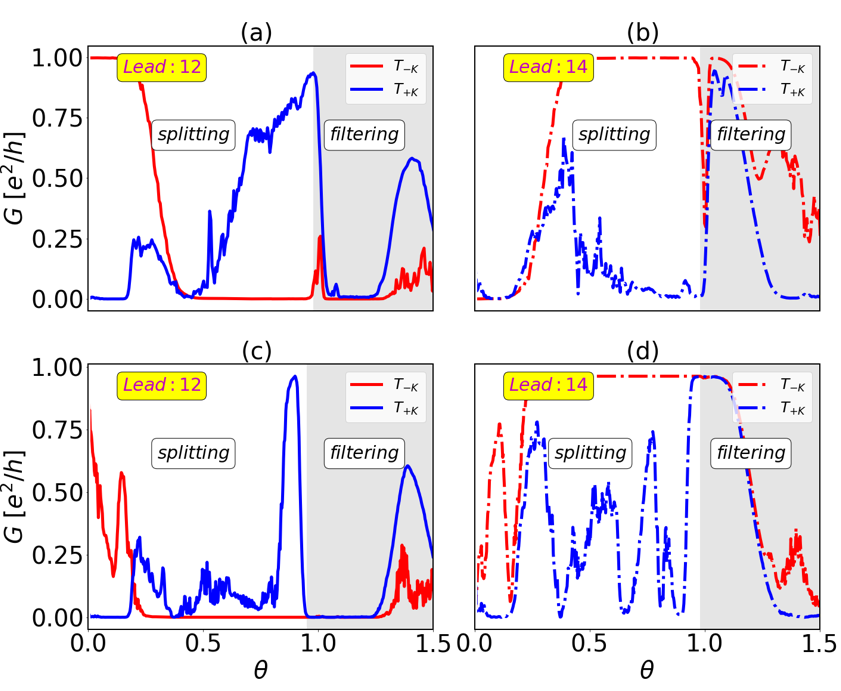}
	\caption{Conductance for each valley as a function of angle $\theta$ (in rad) for $h=8$ nm and $\sigma=4.65$ nm. (a)-(b) occur in the absence of a gate voltage  and  (c)-(d) occur at $V_g=0.5\mathrm{~eV}$. The injection is done through the lead 1.(see Fig.~\ref{fig:T(E)})
	\label{fig:G(teta)}}
\end{figure}

\begin{figure*}
\centering
\begin{subfigure}[b]{0.9\textwidth}		
\includegraphics[width=1\linewidth]{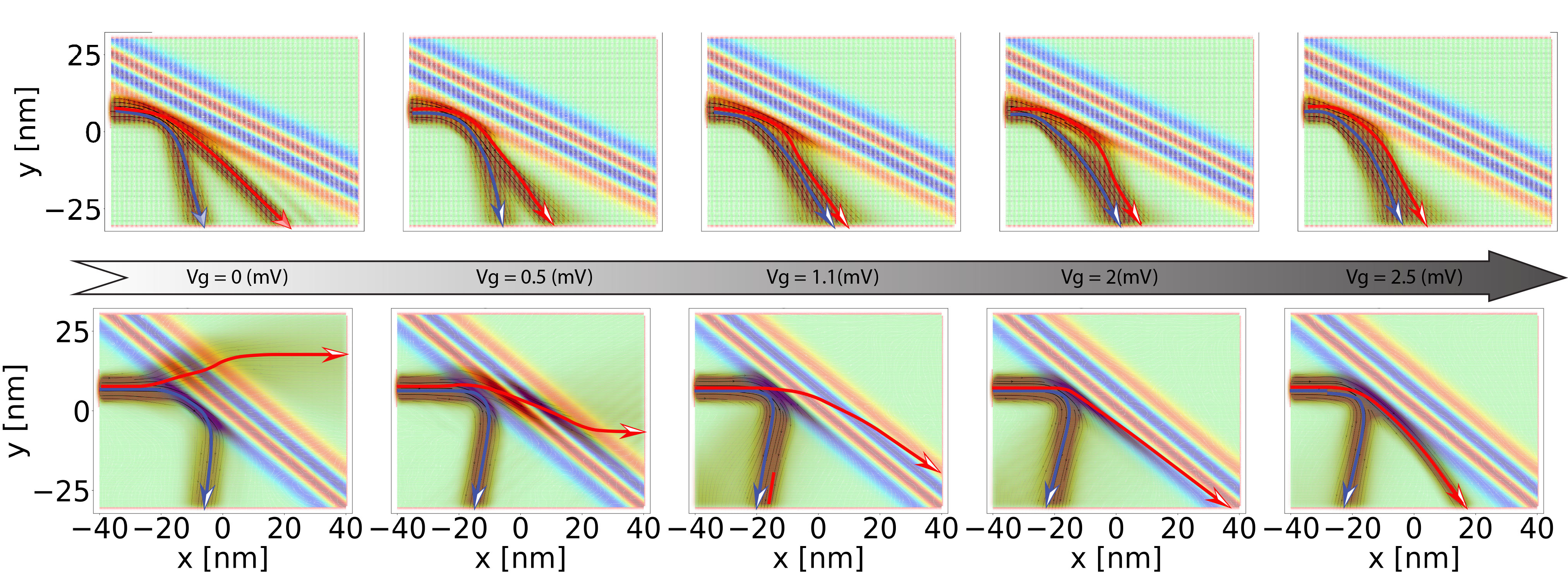}
\end{subfigure}	
\caption{The top row illustrates a situation where valley splitting transitions into valley merging as we increase $V_g$, for deformation parameters $h = 7.95$ nm, $\sigma = 4.65$ nm, and $\theta = 1.10~\mathrm{rad}$. The bottom row shows a transition from current filtering to valley splitting with increasing $V_g$, for $h = 8$ nm, $\sigma = 4.65$ nm, and $\theta = 0.90~\mathrm{rad}$.}
\label{fig7}
\end{figure*}

The angular dependence of the valley-resolved transmission is relatively complex, since several parameters ($\sigma, E, h, \lambda_F, \dots$) play an important role. Nevertheless, from the conductance as a function of $\theta$, we can clearly identify the different transport regimes. Fig.~\ref{fig:G(teta)}(a), for angles slightly smaller than $\theta \approx 1.0~\mathrm{rad}$, only one valley is transmitted to lead 2, while the second valley is suppressed. Conversely, for angles slightly larger than $\theta \approx 1.0 ~\mathrm{rad}$, the first valley becomes suppressed and only the second one is transmitted to lead 4 (see  Fig.~\ref{fig:G(teta)} b). In other words, the two valleys are directed to two different leads, demonstrating a clear valley filtering regime.
When a gate voltage is applied, this picture is slightly modified. In the filtering regime, for example, one of the valleys (blue) is no longer completely suppressed and can be transmitted to both leads 2 and 4, albeit with different amplitudes. It is therefore interesting to investigate how tuning the gate voltage affects the current map for both valleys. Experimentally, this operation is significantly easier than modifying the wrinkle angle.
Fig.~\ref{fig7} illustrates the effect of varying the gate voltage (i.e., the electric field in the $z$-direction) on the current distribution. The top row shows that, starting from a splitting regime, where the two valleys propagate in slightly different directions, increasing the gate voltage progressively reduces the angular separation between them until both valleys propagate in the same direction.
In contrast, the bottom row shows that when starting from a filtering regime, increasing the gate voltage suppresses the transmission of the second valley into lead 2. As a result, both valleys are ultimately transmitted to lead 4. This behavior illustrates a transition from a valley filtering regime to a valley splitting regime.

Gate tuning thus acts as an effective control knob that progressively suppresses valley-selective transport, driving the system from a valley-polarized current distribution to a valley-degenerate regime (Fig.~\ref{fig7} top row). This evolution reflects the role of the gate voltage in enhancing the overlap between valley wave functions, thereby promoting stronger intervalley mixing at higher $V_g$.
Moreover, the transition from a filtering to a splitting regime (Fig.~\ref{fig7}, bottom row) demonstrates that the same device can be reversibly tuned between distinct operational modes through simple electrical gating, without requiring any structural modification.

\section{Conclusion}\label{}
In this work, we have demonstrated that strain-induced wrinkles in graphene nanoribbons provide a simple and effective platform for valley manipulation. The out-of-plane deformation generates localized pseudomagnetic fields that act oppositely on the $K$ and $K'$ valleys, enabling controlled valley splitting and valley filtering without the need for external magnetic fields or complex nanostructuring. Depending on the wrinkle geometry and incidence conditions, the system can operate either in a splitting regime, where the two valleys are spatially separated, or in a filtering regime, where one valley is preferentially transmitted (to a specific lead) while the other is suppressed.

Most importantly, we show that an external electric field applied through a back gate offers dynamic and reversible control over these transport regimes. By tuning the gate voltage while keeping the wrinkle geometry fixed, the device can be driven from valley splitting to valley merging, or from valley filtering to valley splitting. This electrical control provides a practical means to modulate valley polarization and intervalley mixing in a single device.

Our results establish wrinkles as versatile strain-engineered elements for valleytronics and highlight gate-controlled graphene nanostructures as promising candidates for tunable valley-based electronic devices.

\section{Acknowledgement}
A. A, S. E. and M. M. acknowledge the financial support of the Advanced Quantum Computing center at KFUPM.

\begin{widetext}

\end{widetext}

%

\bibliography{references}
\end{document}